\documentclass[12pt, a4paper]{article}
\usepackage{graphicx}
\usepackage[font=footnotesize]{caption}
\usepackage[hscale=0.74,vscale=0.87]{geometry}
\usepackage{amsfonts}
\usepackage{setspace}
\begin{document}

\title{Matter-gravity entanglement entropy and the second law for black holes}

\author{Bernard S. Kay$^*$  \medskip \\  {\small \emph{Department of Mathematics, University of York, York YO10 5DD, UK}} \smallskip \\ 
\small{$^*${\tt bernard.kay@york.ac.uk}}}

\date{}

\maketitle

\centerline{\bf Essay written for the Gravity Research Foundation 2023 Awards for Essays on Gravitation}

\bigskip

\centerline{submitted on 30 March 2023 -- with small corrections 17 May 2023}

\bigskip

\begin{abstract} 
\noindent
Hawking showed that a black hole formed by collapse will emit radiation and eventually disappear.  We address the challenge to define an objective notion of physical entropy which increases throughout this process in a way consistent with unitarity.  We have suggested that  (instead of coarse-grained entropy) physical entropy is matter-gravity entanglement entropy and that this may offer an explanation of entropy increase both for the black hole collapse and evaporation system and also for other closed unitarily evolving systems.   For this to work, the matter-gravity entanglement entropy of the late-time state of black hole evaporation would have to be larger than the entropy of the freshly formed black hole.   We argue that this may possibly be the case due to (usually neglected) photon-graviton interactions.
\end{abstract}

\vfil\eject



\noindent
After the (theoretical) discovery of black holes \cite{LesHouches} and prior to Hawking's (theoretical) discovery \cite{Hawking75} that black holes emit thermal radiation, it had seemed that it would be possible to violate the second law of thermodynamics by simply depositing a package of entropic matter into a black hole.  Bekenstein had pointed out \cite{Bekenstein} that the second law could be saved if a black hole itself had an entropy and argued that this entropy should be proportional to its area.   Hawking's discovery showed that this was indeed the case and (in units with $G=\hbar=c=k=1$) fixed the constant of proportionality to be $1/4$.   Thus an important puzzle was solved.   However Hawking's discovery also led us to question whether we had ever actually properly understood the second law.   One version of the second law is the statement that the entropy of a closed system increases monotonically with time.   And, in a quantum theoretic description based on Hilbert spaces and time-evolving density operators, there is a natural quantity \cite{vonN} with the properties one expects of an entropy, namely the \emph{von Neumann entropy}, $S^{vN}$ (sometimes called the `fine-grained' entropy) defined, for any density operator, $\rho$, to be $-{\rm tr}\rho\log\rho$.   And yet, one must face the fact that the following three assumptions

\begin{itemize}

\item[(1)] Time evolution is unitary

\item[(2)] Physical entropy increases 

\item[(3)] Physical entropy is the von Neumann entropy of the total state

\end{itemize}

\noindent
are in contradiction with one another because von Neumann entropy is a unitary invariant.  We shall call this the \emph{second law puzzle}.

The traditional way to resolve this puzzle \cite{Reich,Davies} is to accept (1) and deny (3) and, instead, to declare that physical entropy is defined by coarse-graining and there are proofs, for specific systems and for suitable initial states, that a suitably defined coarse-grained entropy will then increase, thus satisfying (2).     However, the second law puzzle acquires a new potency when the closed system in question involves a black hole; in particular, for what we shall call here the (gedanken) \emph{black hole collapse and evaporation system} (see Figure 1) which consists initially of a compact star in an otherwise  empty universe which collapses to a black hole which subsequently Hawking radiates and eventually disappears, leaving only Hawking radiation streaming away from where the black hole had been.   For -- see e.g.\ the critique of coarse-graining as applied to this system by Penrose in \cite{Penrose79} -- in view of Hawking's area formula, there seems to be nothing subjective about the entropy of a black hole.

Despite this, a traditional coarse-graining approach has been applied to the black hole collapse and evaporation system with considerable success.   There are indeed proofs (see e.g.\ \cite{Wall2012,Page2013}) that a suitably defined coarse-grained entropy will always increase for this system.   However, motivated partly by the above critique, since 1998 \cite{Kay1998,KayNewt,KayMatGrav}, I have nevertheless explored the prospects for an alternative approach based on a notion of entropy which (unlike coarse-grained entropy) is objective and which, again on the assumption of a unitary dynamics, can again be shown to increase (for suitable initial conditions) both for the black hole evaporation system and also for other systems

\begin{figure}[h]
\centering
\includegraphics[trim = 6cm 19cm 6cm 4cm, clip]{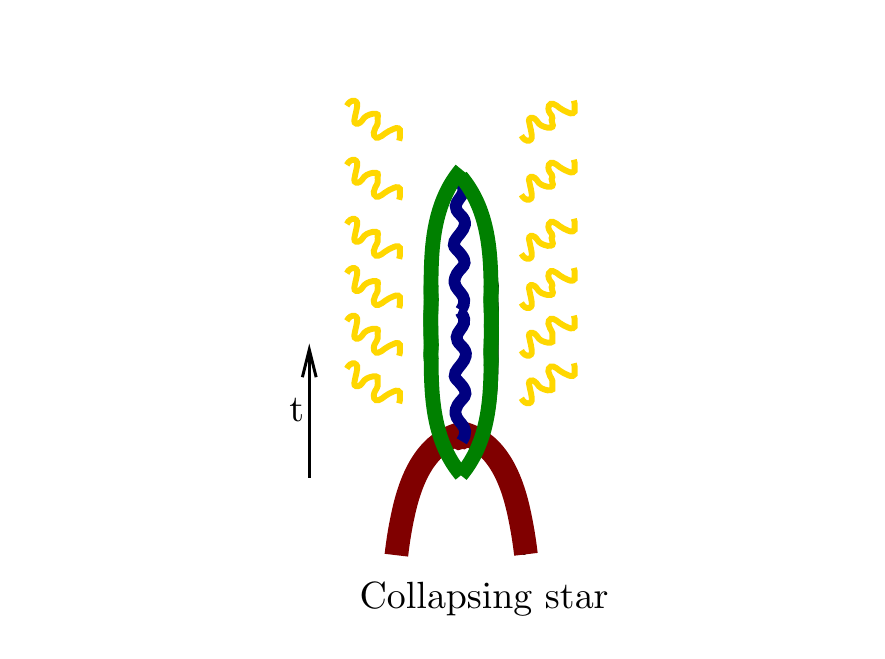}
\caption{A schematic spacetime picture of our (spherical) \emph{black hole collapse and evaporation system}.   The red curves are the boundary of the collapsing star, the green curves the horizon, the black wiggly line the singularity and the yellow squiggles the emitted radiation.}
\end{figure}

Of particular relevance to my train of thought was another puzzle about quantum black holes, the \emph{thermal atmosphere puzzle}.   This concerns not the black hole collapse and evaporation system but rather the equilibrium state that we assume \cite{HawkingBHThermo} to be possible if one confines a (spherical, uncharged) black hole, together with its Hawking radiation, to a spherical box. See Figure 2.   We shall call this the (closed) \emph{black hole equilibrium system}. The puzzle is that there are arguments \cite{PageNewJP,WaldLivRev} that the entropy of such a black hole equilibrium system may be identified with the entropy of the gravitational field of the black hole and there are also arguments \cite{PageNewJP, WaldLivRev} that it may be identified with the entropy of the thermal atmosphere which (in the `inner region' which accounts for most of the entropy -- see below)  consists mostly of matter, albeit a small part of it will consist of gravitons.   And if we were to add those two entropies we would get the wrong answer.  (Twice the correct answer.) 

\begin{figure}[h]
\centering
\includegraphics[scale = 0.7, trim = 6cm 19cm 6cm 5cm, clip]{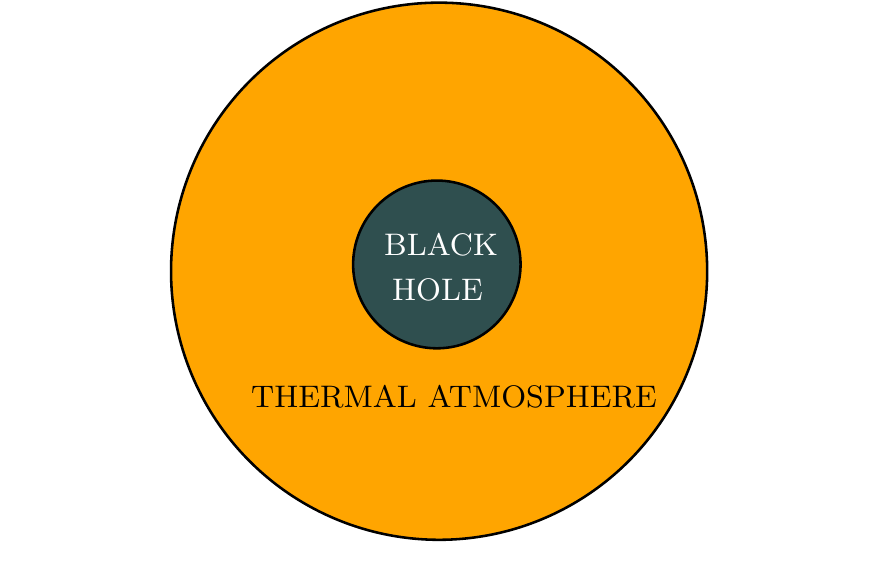}
\caption{A schematic spacetime picture of our \emph{black hole equilibrium system}.}   
\end{figure}

It struck me that this situation is reminiscent of the fact (now very familiar from quantum information theory) that,  if one has a (pure) vector state, $\Psi$, on a bipartite system -- described by a total Hilbert space, $\cal H$, that arises as the tensor product, ${\cal H}_A\otimes{\cal H}_B$ of an `A' Hilbert space and a `B' Hilbert space,  then the partial trace, $\rho_A={\rm Tr}_{{\cal H}_B}(|\Psi\rangle\langle\Psi|)$, of $|\Psi\rangle\langle\Psi|$ over ${\cal H}_B$ (also known as the \emph{reduced state} of the `A' system) is, in general, an impure density operator.   (Similarly with $A \leftrightarrow B$.)  When it is impure, one says that the total state, $\Psi$, is \emph{entangled} and one defines its \emph{A-B entanglement entropy} to be the von Neumann entropy, $S^{\rm{vN}}(\rho_A)=-{\rm tr}(\rho_A\log\rho_A)$ of $\rho_A$ -- which, it turns out, is necessarily equal to the von-Neumann entropy, $S^{\rm{vN}}(\rho_B))$ of $\rho_B$.     One also says that System A \emph{purifies} System B (and vice-versa). 

The obvious connection to try to make is to identify the von Neumann entropy of $\rho_A$ in the above story with `the entropy of the gravitational field of the black hole' and `the von Neumann entropy of $\rho_B$' with `the entropy of the thermal atmosphere' and the A-B entanglement entropy with the entropy of the entire black hole equilibrium system.   Thus it would seem that the thermal atmosphere puzzle would be resolved if we assume the black hole equilibrium system to be correctly described by a pure total state, $\Psi$, of quantum gravity entangled between the gravitational field of the black hole and its atmosphere in such a way that their reduced states are both approximately Gibbs states at the Hawking temperature. And if we identify the physical entropy of the entire closed system with the total state's entanglement entropy between the gravitational field of the black hole and its thermal atmosphere.   (We remark that, previously, the black hole equilibrium system has been modelled by a total Gibbs state at the Hawking temperature or [as in Hawking's discussion \cite{HawkingBHThermo} of black hole thermodynamics] in terms of a total microcanonical ensemble.)   As to evidence that the matter-gravity entanglement entropy takes the Hawking value, one can make a semiqualitative argument \cite{KayThermality,KayMod,KayMore} based on string theory ideas that it goes like the square of the black hole mass (and hence like its area) albeit, along with other string theory approaches to the entropy of Schwarzschild (i.e.\ spherical, uncharged) black holes, one cannot obtain the factor of 1/4.

We want to go further though and to take this as a clue for a universal definition of physical entropy -- i.e.\ a definition for an arbitrary closed system, be it our black hole equilibrium system, be it the black hole collapse and evaporation system, be it the universe!  And we want the notion of physical entropy to make sense, in the case of the black hole collapse and evaporation system, not only during the period of time when there's a black hole present (and emitting radiation) but also before the star collapses and after the black hole has completely evaporated -- and in the case of the universe, at early times soon after the big bang as well as later.  Bearing in mind that, in the case of the black hole equilibrium system, we can roughly equate  `the thermal atmosphere' with `the matter atmosphere' ([near the horizon] gravitons are expected to only be a small part the thermal atmosphere -- we will return to this point later) this suggests that we take our definition for the physical entropy of a closed system to be its \emph{matter-gravity entanglement entropy}. This in turn suggests that we will need as our fundamental framework a low-energy version of quantum gravity (where low energy means below something like the Planck energy, 10$^{19}$ GeV) for presumably `matter' and `gravity' are not fundamental notions in exact quantum gravity but emerge at such `low' energies.  In consequence, we expect that, on the approach to understanding thermodynamics entailed by our proposed definition for physical entropy, thermodynamics is itself an emergent theory that only makes sense at energies below the Planck energy -- and thus only starts to make sense a few Planck times after the big bang when presumably a sort of phase transition occurs and the distinction between matter and gravity starts to emerge. 

On the assumption that low energy quantum gravity is a quantum theory of a conservative type with a total Hilbert space that arises as the tensor product of a gravity Hilbert space and a matter Hilbert space and with a time-evolution, $U(t) = e^{-iHt}$ which is unitary, we then seem to have a chance of resolving the second law puzzle with an objective notion of physical entropy -- namely matter-gravity entanglement entropy -- since this is \emph{not} a unitary invariant.  If we make the additional assumption that the total Hamiltonian, $H_{\rm{total}}$, arises as the sum 
\begin{equation}
\label{Ham}
H_{\rm{total}} = H_{\rm{gravity}} + H_{\rm{matter}} + H_{\rm{interaction}},
\end{equation}
of a gravity Hamiltonian, a matter Hamiltonian and an interaction term, and if we further assume that the initial state (in the case of the universe, a few Planck times after the big bang) is unentangled between matter and gravity (or rather has a low degree of matter-gravity entanglement) then indeed it seems reasonable to expect that the degree of entanglement, and therefore the physical entropy as we define it, will increase because of the interaction.   (If our systems had a finite number of degrees of freedom, one might anticipate Poincar\'e recurrence times when the entropy returned to a low value, but if the number of degrees of freedom is large [or infinite] that might well not happen for a very long time [or ever].)   We shall call this our \emph{general argument for entropy increase}.  Thus our hypotheses and assumptions seem capable of offering a resolution to the second law puzzle by offering a plausible mechanism for why the physical entropy of a closed system increases with time.  At the very least, we can say that with our proposed definition for entropy, the question of whether the second law holds becomes an objective question whose answer depends on the initial state and the (low energy) quantum gravitational Hamiltonian.

We will not pursue that question further in its full generality here.   Rather we shall attempt to  understand, in the case of the black hole collapse and evaporation system, whether (and if so, how) the entropy as we define it, will increase in terms of the details of the collapse and evaporation process (while unitarity is maintained).   Here we remark that, on our matter-gravity entanglement hypothesis, it seems that any configuration of ordinary matter will have a nonzero entropy, albeit this is expected to be small unless the system is highly relativistic and/or involves strong gravitational fields.  (See \cite{KayNewt,KayAbyaneh} where we also give some evidence that the entropy will be lower e.g.\ for a gas and higher when matter is in a more condensed state.)  Thus our collapsing star will already have a small nonzero entropy and presumably this will increase rapidly to the Hawking value as the black hole forms, presumably because of a large degree of matter-gravity entanglement happening around where, in a classical description, a horizon forms.    But also, if the second law is to hold, then it must be that, after the full evaporation of the black hole, the state of Hawking radiation that remains, streaming outwards from the centre where the black hole had been, has a physical entropy even greater than the entropy that the black hole had when it was freshly formed (i.e.\ as we know from Hawking, one quarter of the area of its horizon).  How can that be, if we define physical entropy to be matter-gravity entanglement entropy?   If the initial black hole was large enough and assuming that the only massless particles in nature are photons and gravitons then (cf.\ \cite{Page2013}) the late time Hawking radiation will mainly consist just of photons and gravitons.   For this to have the necessary big entropy as we define it (and for the total state to remain pure!) it presumably must be that the photons and gravitons are highly entangled with one another (while in a total pure state).  (Here we are assuming that matter-gravity entanglement entropy amounts, in this case, to photon-graviton entanglement entropy.)

Now there immediately appears a difficulty.   According to the 1976 Hawking-effect calculations of Page \cite{Page1976, Page1976II}, because of their higher spin, fewer gravitons will be emitted than photons.  Moreover, even if one purifies as many photons as one can with the fewer gravitons predicted to be emitted, one may easily infer from the quantitative results of \cite{Page1976II} (as cited in \cite{Page2013} -- see also \cite{Page1983} and \cite{PageNewJP})  that the resulting matter-gravity entanglement entropy would fall well short of the entropy of the freshly formed black hole.  (In fact, using the results in \cite{Page1976II,Page2013} one finds that one would only obtain $0.15198$ times the entropy of the freshly formed black hole.)  And there is also no mechanism for any such purification in those 1976 calculations.

However there seem to be a couple of important possible loopholes.   Those calculations in \cite{Page1976,Page1976II}, as indeed the original black hole evaporation calculation \cite{Hawking75} of Hawking, are done in the framework of quantum field theory in curved spacetime and it may well be that, at least after a black hole has been radiating for some time, the quantum fluctuations around the horizon build up and render that approximation to quantum gravity (and also semiclassical gravity where one incorporates a backreaction of the expectation value of the stress energy tensor) inapplicable.   Secondly, the work in \cite{Page1976,Page1976II} neglects photon-graviton interactions.   Let us temporarily leave aside our black hole collapse and evaporation system and look in some detail at equilibrium states of quantum gravity confined to a gedanken spherical box of volume $V$.  We model these as  (pure) states chosen at random from superpositions of eigenstates with energies in a narrow band around say $E$.  (See \cite{KayThermality}.)  Although we are dealing with randomly chosen pure total states and Hawking \cite{HawkingBHThermo} studied a microcanonical ensemble, we expect (see \cite{KayThermality,KayMod,KayMore}) that the broad features of the thermodynamical behaviour that is predicted on our assumptions will be qualitatively similar to those described in \cite[Page 195]{HawkingBHThermo}.   (In the terminology of \cite{KayThermality,KayMod} we are replacing the `traditional' by the `modern' explanation of the origin of thermodynamic behaviour.)  If $V$ is sufficiently large, we expect that the box would contain just radiation and no black hole and for that radiation to mainly consist just of massless particles -- i.e.\ photons and gravitons.  But now we would expect the total state to involve equal numbers of photons and gravitons and moreover (on our assumption of a total pure state) for that state to be highly entangled between the photons and gravitons.  After all, all that seems to matter here is that we have a total pure state of two species of massless boson, each with two helicity states and weakly coupled to one another.   And by arguments similar to those in \cite{KayThermality} in such a situation (two weakly coupled systems with the same density of eigenvalues in a pure total state which is a random superposition of energy eigenstates with energies in a narrow band) one expects the state to be highly entangled.

Next consider a suitably smaller volume $V$ (but not so small as for its radius to be within the Schwarzschild radius for energy $E$) then (cf.\ again \cite[Page 195]{HawkingBHThermo}) we would expect a random superposition of energy eigenstates with energies around $E$  to very probably consist of a (spherical) black hole (let us assume, for simplicity, located centrally and ignore its centre of mass motion) surrounded by an atmosphere, thus providing a model for what we called above our `black hole equilibrium system'.   Further, we would expect this atmosphere to roughly consist of an inner region, where $r/2M-1 \ll 1$, where, because of the blueshift, all matter degrees of freedom are effectively massless and (cf.\ 't Hooft's brick wall discussion \cite{tHooft, MukohyamaIsrael} and our discussion of the thermal atmosphere puzzle above) that it is the entanglement of the matter fields in this region with the gravitational degrees of freedom that will account for most of the entropy, and an outer region consisting just of massless particles -- i.e.\ just photons and gravitons.   And, similarly to in the case where a black hole is absent, we would expect those outer photons and gravitons to be in roughly equal numbers and highly entangled with one another, albeit their contribution to the overall (matter-gravity entanglement) entropy will be small compared to that of the inner region.  (It also seems reasonable to expect that the entropy of the inner and outer regions will roughly add up to the total entropy.)  It also seems plausible that those outer photons will (consistently with monogamy \cite{monogamy}) not be strongly entangled with the gravitational field of the inner region.  Now, suppose we were to make a small hole in the box in the latter situation -- or rather, to preserve spherical symmetry, suppose we were to make the box slightly permeable.   Presumably what would come out would be radiation consisting of photons and gravitons --  in roughly equal numbers and highly entangled with one another, while the escaped photons and gravitons would be continually replenished (at the expense of the inner region) as the system inside the box strives to return to equilibrium.  

So, if we start with a black hole in a spherical box in equilibrium and then permit it to evaporate slowly by making the box slightly permeable, it seems possible that what we will be left with after it has fully evaporated, is a pure total state of radiation consisting of photons and gravitons --  in roughly equal numbers and highly entangled with one another.   It seems possible that this might have a (matter-gravity entanglement) entropy greater than that of the freshly formed black hole although whether it does remains an open question still to be investigated.    Let us say in support of this possibility that  it seems possible that the partial state of the photons obtained by tracing over the gravitons (whose von Neumann entropy is of course the photon-graviton entanglement entropy) might not be too different  from the photon component of Hawking radiation calculated using quantum field theory in curved spacetime for a semiclassical model of an evaporating black hole and let us note that, say for an initial Schwarzschild black hole, it easily follows from Page's 1976 calculations \cite{Page1976II,Page1983,Page2013} that the latter would have a (von Neumann) entropy 1.33274 times the entropy of the freshly formed black hole.  

It also seems plausible that the same will be true if the initial state inside the slightly permeable box is not an equilibrium state but the state of a black hole freshly formed by collapse and then enclosed in the box since we would anyway expect that such a system would strive to surround itself with an atmosphere by Hawking radiating.    

Returning to the standard black hole collapse and evaporation system, let us first reiterate that, if we accept our general argument for entropy increase, we would still expect that, at the same low-energy quantum gravity level of description and again due to photon-graviton interactions (which are neglected in a quantum field theory in curved spacetime or semiclassical treatment) the Hawking radiation would still consist, at late times, of a pure state of entangled photons and gravitons with an entropy greater than that of the freshly formed black hole.  To attempt to give a detailed physical understanding of how this might come about seems more difficult and subject to even more uncertainties than for the above-discussed case when the black hole is enclosed by a slightly permeable box.   But we would expect (see the first diagram in Figure 3) that (highly blueshifted) Hawking-emitted photons emerging from just outside the horizon will radiate (soft) gravitons due to the fact that the photons will be freely falling in the gravitational field of (what remains of) the black hole. (The photons will also redshift as they climb out of the gravitational potential.)    One expects such radiation to happen just as one expects that an electron freely falling in the gravitational field of a black hole or of a star will radiate soft photons.\footnote{It has sometimes been argued by an application of the equivalence principle that a charged particle freely falling in a gravitational field will not radiate but there are strong arguments (see e.g.\ \cite{UmGillies}) that such an application of the equivalence principle is illegitimate.  And, anyway (see e.g.\ \cite{MorBerCrisp} and references therein) we know e.g.\ that a charged particle orbiting a black hole will emit synchrotron radiation and also indeed expect that an orbiting massive particle will radiate gravitons.}   Moreover, one expects that as it radiates gravitons, each Hawking photon will become entangled with them, just as \cite{InfraQI,DresInfraQI} (and see \cite{PrabSatWald} for the generalization to massless particles) an outgoing electron in M{\o}ller scattering will radiate, and become entangled with, soft photons.   However, it remains to be investigated whether, in the course of its travel away from the horizon, the resulting degree of entanglement of a photon with its radiated gravitons will be high.  One also, of course, needs to consider what happens to Hawking-emitted gravitons.  One expects (see again Figure 3) Feynman-like diagrams where (in the presence of the gravitational background) such a graviton can produce a pair of photons (at least one of which will be comparably `hard' to the graviton and would then in its turn emit more soft gravitons etc.) and higher order in Newton's constant diagrams where the gravitons continue on their way after emitting even numbers of soft photons as in the third diagram of Figure 3.  Again it remains to be investigated whether the resulting degree of graviton-photon entanglement is high.

\begin{figure}[h]
\centering
\includegraphics[scale=0.20, bb=-2in 2in 14in 10in, clip]{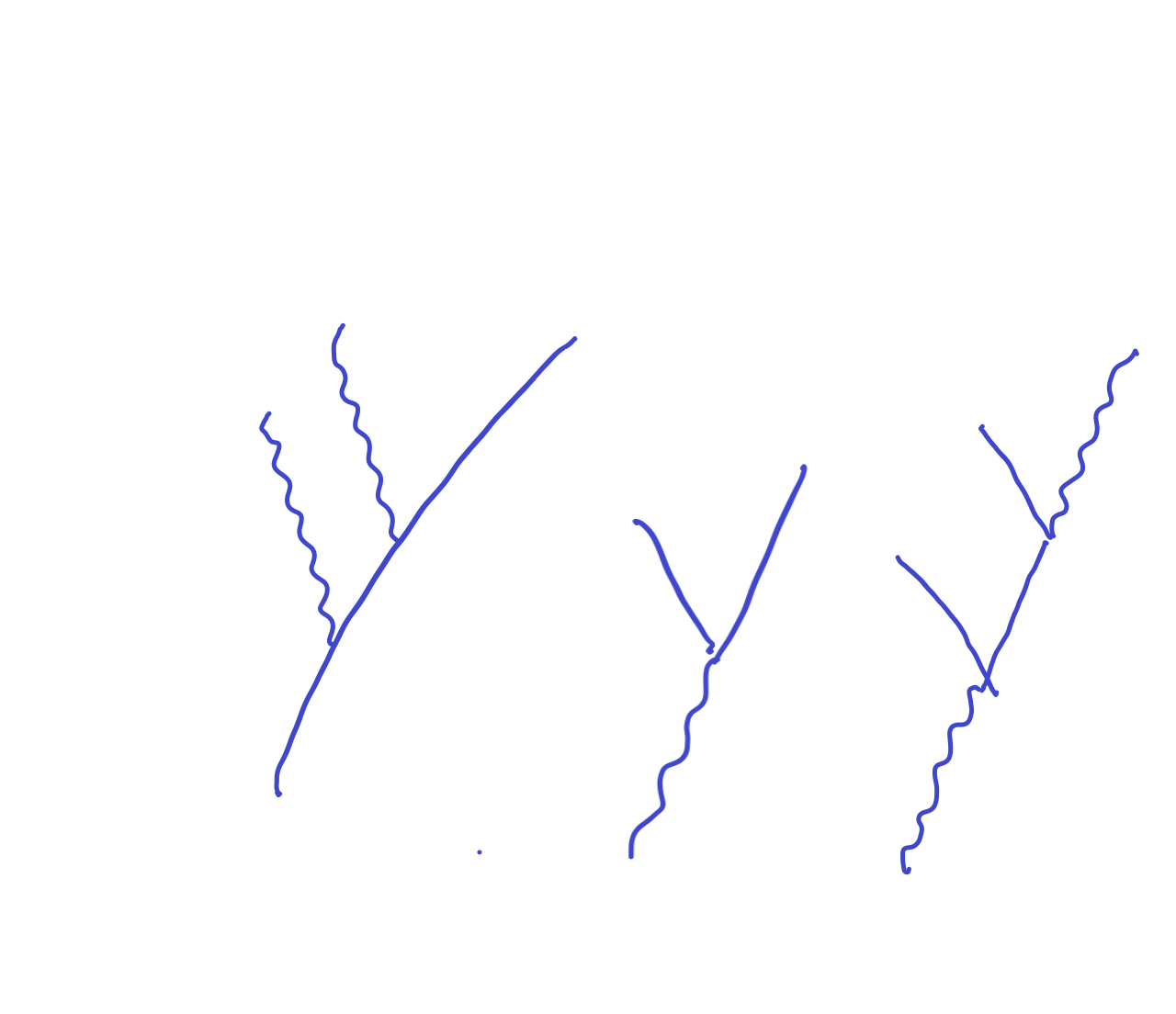}
\caption{Feynman-like diagrams (straight line $=$ photon, wiggly line $=$ graviton) showing a photon emerging from near the horizon emitting (soft) gravitons; a (hard) graviton emitting (say) a hard and a soft photon; and a graviton converting to a photon and then back to a graviton after emitting two (soft) photons}
\end{figure}

In conclusion, we have sketched possible mechanisms based on photon-graviton interactions (and on interactions of other matter particles with gravitons) for how entropy as we define it  (i.e.\ matter-gravity entanglement entropy) might increase throughout the formation and subsequent evaporation of a black hole, thereby also offering support for our above general argument for entropy increase in the case of our black hole collapse and evaporation system.  If the freshly formed black hole is enclosed in a slightly permeable box,  we argued that this is possible by relating the late-time state of Hawking radiation to the state of the outer part of the thermal atmosphere in the black hole equilibrium system.   In the absence of such a box, we pointed out that one would expect i.a.\ Hawking-emitted photons themselves to radiate gravitons and become entangled with them.  More work is needed to test if these possibilities actually hold, in particular whether the latter degree of entanglement will be high enough for the entropy to always increase.   If it will be, then one would expect the late-time state of Hawking radiation to largely consist of photons highly entangled with, and purified by, gravitons.

\end{document}